\documentclass[aps,amsmath,graphicx,twocolumn]{revtex4}
\usepackage{graphics}
\usepackage{graphicx}
\usepackage{amsmath}
\usepackage{amssymb}
\usepackage{epsfig}
\begin{document}

\title{Hyperfine field assessment of the magnetic structure of ZrZn$_2$}

\author{A.V. Tsvyashchenko}
\email{tsvyash@hppi.troitsk.ru}
\affiliation{Vereshchagin Institute for High Pressure Physics, RAS, 142190, Moscow, Troitsk, Russia}
\affiliation{Skobeltsyn Institute of Nuclear Physics Lomonosov Moscow State University, Leninskie gory, Moscow 119991, Russia}

\author{D.A. Salamatin}
\affiliation{Vereshchagin Institute for High Pressure Physics, RAS, 142190, Moscow, Troitsk, Russia}
\affiliation{Moscow Institute of Physics and Technology, 141700 Dolgoprudny, Russia}

\author{A. Velichkov}
\affiliation{Joint Institute for Nuclear Research, Dubna, P.O. Box 79, Moscow, Russia}

\author{A.V. Salamatin}
\affiliation{Joint Institute for Nuclear Research, Dubna, P.O. Box 79, Moscow, Russia}

\author{L.N. Fomicheva}
\affiliation{Vereshchagin Institute for High Pressure Physics, RAS, 142190, Moscow, Troitsk, Russia}

\author{V.A. Sidorov}
\affiliation{Vereshchagin Institute for High Pressure Physics, RAS, 142190, Moscow, Troitsk, Russia}

\author{A.V. Nikolaev}
\affiliation{Skobeltsyn Institute of Nuclear Physics Lomonosov Moscow State University, Leninskie gory, Moscow 119991, Russia}
\affiliation{Moscow Institute of Physics and Technology, 141700 Dolgoprudny, Russia}

\author{A.V. Fedorov}
\affiliation{Vereshchagin Institute for High Pressure Physics, RAS, 142190, Moscow, Troitsk, Russia}

\author{G.K.  Ryasny}
\affiliation{Skobeltsyn Institute of Nuclear Physics Lomonosov Moscow State University, Leninskie gory, Moscow 119991, Russia}

\author{V.N. Trofimov}
\affiliation{Joint Institute for Nuclear Research, Dubna, P.O. Box 79, Moscow, Russia}

\author{A.V. Spasskiy}
\affiliation{Skobeltsyn Institute of Nuclear Physics Lomonosov Moscow State University, Leninskie gory, Moscow 119991, Russia}

\author{M. Budzynski}
\affiliation{Institute of Physics, M. Curie-Sklodowska University, 20-031 Lublin, Poland}

\begin{abstract}
Time differential perturbed angular $\gamma \gamma$-correlation (TDPAC) method on $^{111}$Cd nuclei probes inserted in ZrZn$_{1.9}$ is used to measure the magnetic hyperfine
fields (MHF’s) at Zr and Zn sites and the electric field gradient (EFG) $V_{zz}$  at Zn sites as a function of temperature at various pressures and as a function of pressure at the
temperature 4 K. Our data indicate that the local magnetic moment of Zr in the magnetically ordered state is substantially larger than its value obtained from the macroscopic
measurements and that there is also an induced magnetic moment at the Zn site. We conclude that ZrZn$_2$ is not a simple ferromagnet and discuss a possible type of its magnetic
ordering.
\end{abstract}
\maketitle

\section{INTRODUCTION}

Itinerant ferromagnetism attracts substantial interest despite its being relatively less common. This interest is largely triggered by interesting and not yet fully understood phenomena
that occur when the order is suppressed at zero temperature by pressure or other means. Such quantum phase transitions (QPT) are driven by quantum fluctuations rather than
thermal ones, and the transition itself represents a quantum critical point (QCP). In the vicinity of the quantum critical point novel and nontrivial phenomena emerge, such as non-
Fermi liquid behavior\cite{1}, triplet superconductivity\cite{2,3,4,5,6}, skyrmion phases and the topological Hall effect\cite{7,8}.

ZrZn$_2$ is a prototypical itinerant ferromagnet, and it exhibits a first order QPT at the critical pressure $p_c$ = 16.5 kbar, in accord with the metamagnetic behavior, characterized
by a sudden superlinear rise in the magnetization as a function of applied field, for pressures above $p_c$ \cite{9}.

ZrZn$_2$ crystallizes in the C15 cubic Laves structure. At ambient pressure the ferromagnetic order sets in at the Curie temperature $T_C$ = 23--28 K. Magnetization
measurements indicate a small net magnetic moment of 0.13--0.23 $\mu_B$ per formula unit, while the intermediate-temperature susceptibility can be fitted by the Curie-Weiss law
with the effective moment 1.9 $\mu_B$ \cite{9,10}  Not unusual for itinerant magnets, the overall temperature dependence is not Curie-Weiss; as the temperature grows above the
room temperature, the effective moment is reduced and the Curie-Weiss temperature increases compared to $T_C$. Unlike UGe$_2$ \cite{11} and MnSi, \cite{12,13,14} the
magnetic structure of ZrZn$_2$ has not been successfully characterized by neutron diffraction. Neutron scattering on high-quality samples of ZrZn$_2$ have provided the only
direct experimental evidence on the nature of the spin density\cite{15,16}  and low-lying magnetic excitations\cite{17}. They have confirmed the itinerant nature of spin polarization,
as the latter is delocalized along the network of Zr atoms with the maximum polarization at the middle of Z-Zr bonds.  The existence of a magnetic short-range order in ZrZn$_2$
however has been confirmed by nuclear magnetic resonance (NMR) on the $^{91}$Zr nuclei \cite{18,19,20}. In addition, a local magnetic hyperfine field (MHF) at Zr site was
reported in a few studies \cite{21,22}  allowing to estimate the value of its magnetic moment.


While it is generally accepted that ZrZn$_2$ is a uniform ferromagnet with all Zr atoms carrying the same (small) magnetization, existing neutron scattering experiments\cite{15,16}
were primarily aimed at distinguishing between the localized and itinerant magnetism, and cannot exclude with certainty a possibility of a sign change, for instance, a long-pitch
spiral.

Remarkably, a magnetic component of the opposite sign, that is, the antiferromagnetic component has been unambiguously detected in the ZrZn$_2$ powder sample in the zero-
field muon spin rotation ($\mu$SR) experiments\cite{23}. The $\mu$SR data imply that the magnetically ordered phase of ZrZn$_2$ is not a simple ferromagnet.

It is also worth noting that the standard density functional theory (DFT) in the local density approximation (LDA) results in a large magnetic moment\cite{24} of 0.72 $\mu_B$ per
formula unit (f.u.) for ZrZn$_2$. This value is nearly four times larger than the established experimental moment. In Ref. \onlinecite{24} this fact is accounted for by fast spin
fluctuations averaged with the help of Moria’s self-consistent renormalization procedure.

In the present work we report the magnetic hyperfine fields at Zr and Zn sites and the electric field gradient (EFG) at Zn sites of ZrZn$_{1.9}$ measured by means of time-differential
perturbed angular correlations (TDPAC) spectroscopy at probe $^{111}$Cd nucleus which has a close electron shell in the lattice. The probe $^{111}$Cd nuclei were inserted in
the ZrZn$_{1.9}$ lattice and were detected at both Zr and Zn sites of ZrZn$_{1.9}$ crystallized in the C15 cubic Laves structure (as ZrZn$_2$). The experiments have been carried out
at low temperatures with applied pressure up to 2 GPa.

Note that the hyperfine spectroscopic techniques such as TDPAC, M\"{o}ssbauer, NMR and others are very sensitive to inhomogeneities of the magnetic properties, which is their very
important advantage. In particular, TDPAC spectroscopy measuring short ranged hyperfine interactions (magnetic or electric) results in a nanometer special resolution and can
identify different local configurations in the same sample. Unlike macroscopic techniques that measure averaged quantities such as magnetization, resistivity etc., TDPAC is a
microscopic method that can determine local variations of magnetic moment and exchange.

The paper is organized as follows. In Sec. 2 we give experimental details of the TDPAC measurements, in Sec. 3 we analyze our data on ZrZn$_{1.9}$ and in Sec. 4 present the
summary of our work.

\section{EXPERIMENTAL}

Using the TDPAC technique, as described below, we have measured both the magnetic hyperfine fields (MHF) and the EFG introducing well-known nuclear probes $^{111}$In/$^
{111}$Cd at Zr- and Zn-sites of the ZrZn$_{1.9}$ polycrystalline sample synthesized at high pressure. Earlier we have demonstrated that our measured spectra of angle anisotropy
are more refined \cite{25} (We observed 100\% absorption of the introduced $^{111}$In/$^{111}$Cd impurities in the crystal lattice of ZrZn$_{1.9}$). The synthesized samples of for
that composition (i.e. ZrZn$_{1.9}$) are characterized by the highest and most reproducible transition temperature \cite{26} (i.e. $T_c$ is reproducible up to ±0.5 K) among all
considered non-stehiometric compounds.

The parent isotope $^{111}$In with high specific activity was obtained using the $^{109}$Ag($\alpha$, 2n) $^{111}$In reaction by irradiating a silver foil in the 32 MeV $\alpha$-beam at
the Nuclear Physics Institute cyclotron (Moscow State University). The $^{111}$In has a long enough half-life $T_{1/2}$=2.83 d which permits carrying out experiments during up
to two weeks using one portion of the initial activity. After the electron capture decay of $^{111}$In, $^{111}$Cd is formed in the 420 keV excited state, which de-excites by the $
\gamma$-ray cascade 173–-247 keV. The intermediate 247 keV state has the spin $I=5/2$, electric quadrupole moment $Q$ = 0.83 b and $T_{1/2}$ = 85 ns.
Small pieces of the irradiated foil of Ag ($m\, <$ 1 mg) have been melted together with powdered Zr and Zn  (with chemical purity of 99.99\% and 99,999\%, respectively,  and a total
mass 500 mg) in ratio 1:1.9 at a pressure of 8 GPa.  The ingots were crushed and small bright fragments from the inner parts of the ingots were used for the TDPAC experiments.
After the $^{111}$In activity had practically decayed out, the x-ray diffraction off the samples was measured. All samples had a pure cubic C15 Laves phase structure.

Ferromagnetic ZrZn$_2$ is crystallized in the cubic C15 lattice structure with Zr and Zn forming two sublattices. All sites within each of the sublattices are equivalent. The local
symmetry of the Zr site is tetrahedral (the $T_h$ site symmetry), and that of the Zn site is noncubic (the $3m$ site symmetry). This difference in symmetry is instrumental in
assigning the probes among two sublattices. While the tensor of the electric-field gradient (EFG) is nonzero at the Zn sites, it vanishes at all Zr sites. Consequently, at $T > T_c$ a
finite electric quadrupole interaction (QI) is expected for $^{111}$Cd at the Zn site and zero at the Zr site. Thus, the $^{111}$Cd probe at the Zn site experiences both the electric
field gradient (EFG) and the magnetic hyperfine field (MHF), which couple to the nuclear electric quadrupole ($Q$ = 0.83 b) and the magnetic dipole ($\mu$) moment of the
intermediate nuclear state, respectively. In the proper reference frame (with the diagonal components of the tensor $V_{ij}$ of EFG) the Hamiltonian for such static interactions
reads:
\begin{equation}
 H = \frac{\hbar \omega_0}{6} [3I_z^2 - I(I+1) + \frac{1}{2}\eta (I_+^2 + I_-^2) ]
 + \vec{\mu} \vec{B}_{hf} .
\end{equation}
Here $\omega_0=3eQV_{zz}$ / [$2I(2I-1)\hbar$] is the fundamental precession frequency, $I$ represents the nuclear spin of the probe intermediate state ($I=5/2$ for
$^{111}$Cd), the asymmetry parameter $\eta = (V_{xx}-V_{yy})/V_{zz}$, $V_{ii} = \partial^2 V/\partial^2 i $ ($i=x, y, z$) are the principal-axis components ($|V_{zz}| \ge |V_
{yy}| \ge |V_{xx}|$) of the EFG tensor, $\vec{B}_{hf}$ is the magnetic hyperfine field. Finally, $\omega_L = 2\pi \nu_L = -g \mu_N B_{hf} / \hbar$ is the Larmor frequency and
the $g$-factor of the $I=5/2$ state of $^{111}$Cd is \cite{27} $g$ = $-$0.306. The time evolution of the perturbed $\gamma-\gamma$ correlation is described by the experimental
function $R(t)$, where $t$ is the time spent by the nucleus in the $^{111}$Cd intermediate state. For a hyperfine interaction, $R(t)$ may be expanded as
\begin{equation}
  R(t) = \sum A_{kk} G_{kk}(t) ,
\end{equation}
where $A_{kk}$ are the angular correlation coefficients. The perturbation factor $G_{kk}(t)$ is a signature of the fields interacting with the probes. These are MHF and an EFG in
the ferromagnetic phase ($T < T_c$) and EFG alone for $T > T_c$. Thus, below $T_c$ in we took into account both interactions to obtain the MHF and EFG parameters. Above
$T_c$, on the other hand, considering only electric quadrupole interaction, the perturbation factor $G_{kk}(t)$ is expressed as \cite{27}
\begin{equation}
  G_{kk}(t) = S_{k_0} + \sum_n S_{k_n} \cos(\omega_n t) e^{-\omega_n \delta t} .
\end{equation}
The frequencies $\omega_n$ and amplitudes $S_{k_n}$ are determined from the diagonalization of $H$. For spin $I=5/2$, three frequencies are observable that are functions of
$\omega_0$ and $\eta$ \cite{28}. Here, we restrict ourselves by the perturbation parameter of the second order since the unperturbed angular correlation coefficient $A_{44} \ll A_{22}$
($A_{22}$ = --0.18).


The perturbation factor $G_{22}(t) = \sum_i f_i G_{22}^i(t)$ (where $f_i$ are partial populations of different sites) describing a nuclear spin precession due to a hyperfine
interaction was determined in a usual way from the angular anisotropy $R(t)$:
\begin{equation}
  R(t) = -A_{22} Q_2 G_{22}(t) .
\end{equation}
$R(t)$ was obtained by combining the delayed coincidence spectra $N(90^{\circ},t)$ and $N(180^{\circ},t)$ measured at the angles of  90° and 180° between detectors, respectively:
\begin{equation}
  R(t) = -2 [ N(180^{\circ},t) - N(90^{\circ},t) ] / [ N(180^{\circ},t) + 2N(90^{\circ},t) ] .
\end{equation}
Here $Q_2\, \approx \, 0.80$ is the solid-angle correction.

The TDPAC measurements were carried out using a four detector spectrometer \cite{29} equipped with an optical four-window cryostat “JANIS” (model SHI-950). The modified
channel of cryostat had the high-pressure chamber of piston-cylinder type \cite{30}, capable of generating a sample pressure of 2 GPa. The hyperfine interaction parameters were
extracted from the measured perturbation functions $A_{22}G_{22}(t)$ using the DEPACK program developed by Lindgren\cite{31}.
Magnetic ac-susceptibility was measured in piston-cylinder type pressure chamber\cite{30} in the sample which was used for TDPAC measurements.

\section{RESULTS AND DISCUSSION}

The interactions of the magnetic dipole moment of the $I=5/2$, 247 keV state of $^{111}$Cd with the magnetic hyperfine fields acting at the probe sites of Zr and Zn of  ZrZn$_{1.9}$
were determined from the time dependence of the anisotropy $R(t)$ of $^{111}$Cd at different temperatures. Fig. 1 illustrates the typical thermal evolution of the TDPAC spectra
(Notice that the transition temperature is $T_C$ = 23 K).
\begin{figure}
\resizebox{0.3\textwidth}{!}
{
 \includegraphics{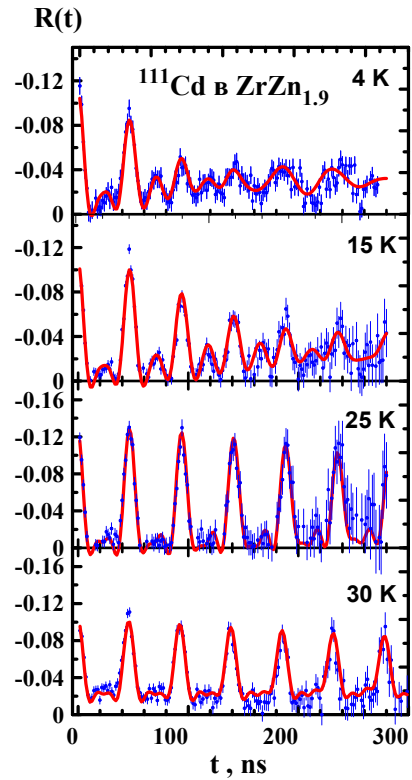}
}
\caption{
Time spectra of the angular correlation anisotropy, $R(t)$ for $^{111}$Cd in ZrZn$_{1.9}$ measured at various temperatures and normal pressure ($T_C$ = 23 K).
}
\label{fig1}
\end{figure}

The $^{111}$Cd -TDPAC spectrum measured at normal pressures and the temperature of 4 K indicates that 30\% of the $^{111}$Cd nuclear probes are located at the Zr sites,
while 70\% occupy the Zn sites. The probe at the Zr site having the tetrahedral symmetry (cubic site i.e. no EFG) is described by a single magnetic frequency $\nu_{L,Zr}$ = 21.3
(3) MHz ($B_{hf} = 2 \pi \nu_L \hbar/g\mu_N$, $B_{Zr}$ = 9.2 T). The probe at the Zn site in addition is experiencing the nuclear quadrupolar interaction with EFG and the
induced magnetic hyperfine field (IMHF). The quadrupolar frequency $\nu_{Q,Zn}$  is 141(1)  MHz with $\eta$ = 0 (which corresponds to the EFG $V_{zz} = \nu_Q h/eQ$ =
7.2(2) 10$^{17}$ V/sm$^2$), and the induced magnetic frequency $\nu_{L,Zn}$ = 2.5(3) MHz
($B_{Zn}$ = 1.2 T). The quadrupolar frequency $\nu_{Q,Zn}$  was practically
the same in the whole temperature range at normal pressure and high pressures.

From the fitting analysis we also extract information concerning angles $\beta$ and $\gamma$ giving the direction of the magnetic hyperfine field in the interaction coordinate system
defined by the EFG tensor. Since for the C15 Laves phase the principal axis of $V_{zz}$ coincides with the crystallographic [111] axis, these angles define the direction of the
IMHF with respect to the [111] axis. We have obtained $\beta$ = 51(5) grad and $\gamma$ = 0, indicating that the direction of the induced magnetic field approximately coincides
with the crystallographic [100] axis, consistent with the equivalence of all magnetic Zn sites in the TDPAC measurements.


Earlier, in ZrZn$_2$ the magnetic hyperfine field (MHF) of $-1.7$ T was reported at 4.2 K \cite{21} measured with the TDPAC spectroscopy at probe $^{181}$Ta nuclei placed at
zirconium sites. However, probe nuclei with open electron shell (like Ta) are not a good choice, because they substantially modify the local $d-s$ polarization of zirconium valence
electrons and inevitably distort the resultant magnetic hyperfine field.

Note also that the values of MHF determined by NMR and TDPAC spectroscopy at the $^{91}$Zr nucleus and the $^{111}$Cd probe nucleus, correspondingly, are different.
Indeed, in the paramagnetic temperature range of ZrZn$_2$ NMR can determine only the coupling constant $A(4d) = H_{hf}(d)/\mu_B$, describing the $4d-$contribution. (Here
$A(4d)$ is found from Knight shifts $K_d(T)=H_{hf}(d) \chi_d(T)/\mu_B$, and the magnetic susceptibility of $d$-electrons 
$\chi_d(T)= \chi(T) - \chi_{dia}  - \chi_{orb} - 2/3\chi_s - 2\chi^{Zn}$ 
for $^{91}$ZrZn$_2$, Ref. \onlinecite{20}). In order to find $A(4d)$ in the ferromagnetic temperature range one has to know the dependence of $A(4d)$
on external magnetic field and its value at zero magnetic field (see details in Ref. 20). The extracted value of $A(4d)$ is approximately an order of magnitude smaller than the $4d$
constants for Rh and Pd \cite{20}. The authors attributed the low value of $A(4d)$ to the negative contribution to MHF from the polarization of $s$-electrons as a result of the $s-
d$  hybridization and the $s-d$ exchange. These contributions are different in the case of $^{111}$Cd probe in the TDPAC measurements, because $^{111}$Cd probes have a
close $s$-shell which is less susceptible to the polarization at the probe atom. Therefore, the values of MHF measured at $^{91}$Zr nuclei in the NMR method differ from the
values of MHF measured at $^{111}$Cd probe nuclei in the TDPAC spectroscopy.
In general, TDPAC has the advantage over zero field NMR that it does not require a frequency sweep to explore the distribution of MHF and has greater sensitivity and resolution
than the M\"{o}ssbauer effect, or $\gamma$-ray asymmetries (method of nuclear orientation).

The temperature dependence of the MHF for $^{111}$Cd at the Zr site and at the Zn site is shown in Fig. 2. While the MHF $B_{Zr}$ remains constant up to the Curie temperature
$T_C$, $B_{Zn}$ demonstrates a remarkable decrease, and disappears at $T_C$. This is accompanied by a decrease in the number of magnetic $^{111}$Cd probes at the Zr
sites, whereas the number of magnetic probes at the Zn site remains the same. Thus, the temperature dependence of $B_{Zr}$ shown in Fig. 2 is typical for the evolution of an
order parameter during a first-order phase transition \cite{32}.
\begin{figure}
\resizebox{0.4\textwidth}{!}
{
 \includegraphics{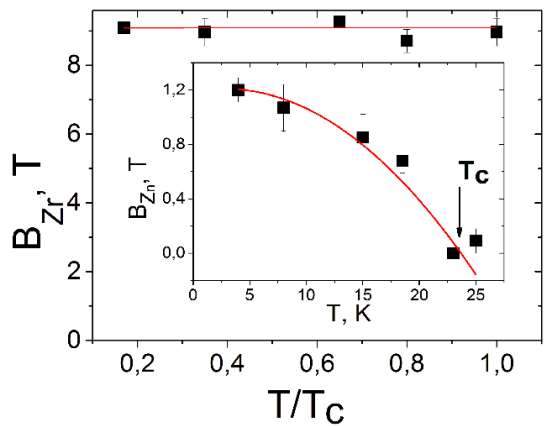}
}
\caption{
Temperature dependences of the magnetic hyperfine fields $B_{Zr}$ and $B_{Zn}$
($B_{hf} = 2 \pi \nu_L \hbar/g \mu_N$) at probe $^{111}$Cd nuclei at Zr sites and Zn sites in ZrZn$_{1.9}$.
}
\label{fig2}
\end{figure}

The pressure evolution of the TDPAC spectra measured at $T$ = 4 K is shown in Fig. 3. At the pressure $P$ = 0.78 GPa and above we found that $\nu_{L,Zr} = 0$ and $\nu_
{L,Zn} = 0$. The dependencies of the Larmor frequencies $\nu_{L,Zr}$ and $\nu_{L,Zn}$ when 
$P\, < \, 0.78$ GPa are plotted in Fig.~4. While $\nu_{L,Zr}$ does not change up to
0.6 GPa, $\nu_{L,Zn}$ decreases linearly. At a pressure above 0.6 GPa the Larmor frequencies $\nu_{L,Zr}$ and $\nu_{L,Zn}$ drop discontinuously. This indicates that the
IMHF $B_{Zn}$ depends on interatomic distances, while $B_{Zr}$ may be responsible for the first order quantum phase transition \cite{8} observed in ZrZn$_2$ at a high
pressure.  The linear pressure dependence of $\nu_{L,Zn}$ is very similar to that of the ordered magnetic moment $M$ in Ref. \onlinecite{8}, which was obtained by extrapolating
magnetic isotherms (Arrott plots) to zero field. This observation suggests that the main contribution to the magnetization of ZrZn$_2$ is caused by the magnetic moment (and
correspondingly by IMHF $B_{Zn}$) induced at the Zn site.
\begin{figure}
\resizebox{0.3\textwidth}{!}
{
 \includegraphics{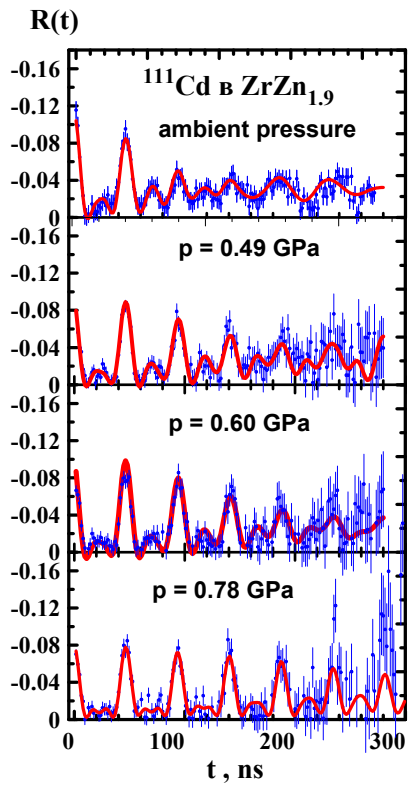}
}
\caption{
Time spectra of the angular correlation anisotropy, $R(t)$ for $^{111}$Cd in ZrZn$_{1.9}$ measured at various pressures and  the temperature T = 4 K.
}
\label{fig3}
\end{figure}
\begin{figure}
\resizebox{0.4\textwidth}{!}
{
 \includegraphics{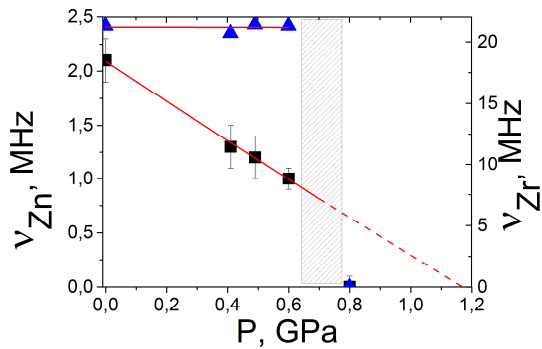}
}
\caption{
Pressure dependences of the magnetic frequencies $\nu_{Zr}$  and $\nu_{Zn}$ measured at the temperature $T$ = 4 K for the probe $^{111}$Cd nucleus at the Zr site and
the Zn site in ZrZn$_{1.9}$. ($\blacksquare$) data for the Zn site ($\nu_{Zn}$) and ($\blacktriangle$) data for the Zr site ($\nu_{Zr}$).
}
\label{fig4}
\end{figure}

In \cite{16} it has been demonstrated that the $4d$ and $5p$ contributions to magnetization at the Zr site amounts to 57\% while the remaining part of magnetization (43\%)
does not lead to Bragg scattering at finite angles. The authors explain that by a contribution to magnetization which is not connected with the Zr band electrons. This however does
not exclude hybridization between $4d$ states of Zr and polarized 4p states of Zn. The hybridization leads to induced magnetization at the Zn site and results in induced magnetic
hyperfine field (IMHF) $B_{Zn}$. Notice also that in Refs. \onlinecite{15,16}  spin density was detected not only at Zr sites but also at midway positions of the Zr atoms at $z =0$
(1/2) and at $z = ј$ (3/4). This indicates that four tetrahedra formed by Zn atoms around a Zr site experience a homogeneously distributed magnetization from hybridised Zr $4d$
($T_{2g}$) : Zr $5p$ electron states \cite{16}, which also leads to IMHF $B_{Zn}$ at the Zn site.
Fig. 5 illustrates the pressure evolution of the Curie temperature $T_C$ for the same sample of ZrZn$_{1.9}$ measured by $^{111}$Cd-TDPAC technique and magnetic ac-
susceptibility. One can clearly see a good correspondence between these two methods. The difference in the values of $T_C$ can be explained by the fact that the microscopic
(TDPAC) method has a higher sensitivity to the local changes of magnetic interactions.
\begin{figure}
\resizebox{0.4\textwidth}{!}
{
 \includegraphics{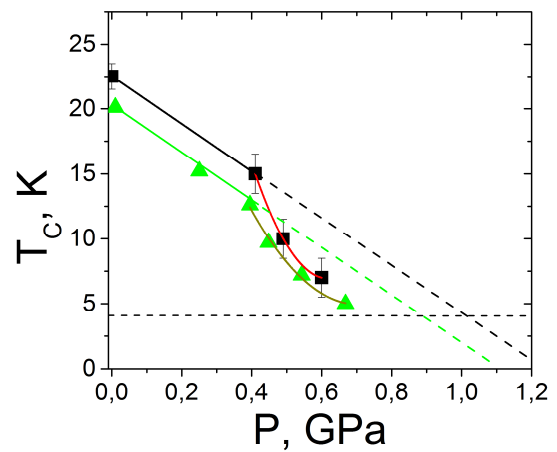}
}
\caption{
Pressure dependences of Curie temperature $T_C$ for the same sample of ZrZn$_{1.9}$  ($\blacksquare$) data of the $^{111}$Cd-TDPAC measurements
and ($\blacktriangle$) the magnetic ac-susceptibility measurements
(see also Fig. 6).
}
\label{fig5}
\end{figure}

Fig. 6 shows the temperature dependence of the magnetic ac-susceptibility of ZrZn$_{1.9}$ measured at pressures up to 1.61 GPa and temperatures down to 4.2 K. (Here we have
used of a miniature chamber of the piston-cylinder type \cite{30}). As pressure increases up to 0.4 GPa the Curie temperature $T_C$ of ZrZn$_{1.9}$ reduces linearly. For pressures
above 0.4 GPa the Curie temperature $T_C$ exhibits a nonlinear dependence with a sudden drop. It has been demonstrated previously that polycrystalline samples of ZrZn$_2$
synthesized at high pressure, show basically the pressure evolution reported for ZrZn$_2$ earlier albeit they are also highly sensitive to the quality and the history of samples 
\cite{33}. 
Therefore, the deviation from the linear decrease of $T_C$ in ZrZn$_{1.9}$ is most likely caused by vacancies which lead to a decrease of the exchange interaction.
\begin{figure}
\vspace{0cm}
\resizebox{0.4\textwidth}{!}
{
 \includegraphics{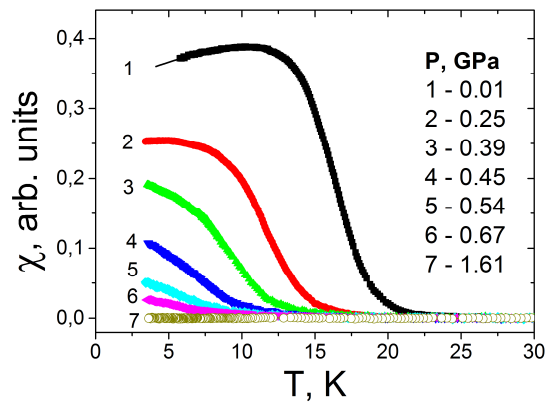}
}
\caption{
Temperature dependences of the magnetic ac-susceptibility of ZrZn$_{1.9}$ measured at pressures up to 1.61 GPa and temperatures down to 4.2 K.
}
\label{fig6}
\end{figure}

{\it Ab initio} calculations of the hyperfine field on Cd impurities with scaled exchange-correlation potential (as in Ref. \onlinecite{34}) yield the coupling constants $A_{Zr}$ and
$A_{Zn}$ for Cd(Zr) and Cd(Zn) impurities ($A_{Zr}$ = -18.4 $T/\mu_B$ and $A_{Zn}$ = -4.2 $T/\mu_B$) \cite{35}.
It is worth noting that earlier the magnetic hyperfine field $B_{hf}$ was found to be approximately proportional to the inducing $d$ moment in the $3d$ hosts Fe, Co, Ni \cite{36,
37}: for the closed-shell nucleus $^{111}$Cd, e.g. $B_{hf}$ was a linear function of $\mu_{3d}$ with the coupling constant $A_{3d} = B_{hf} /\mu_{3d}$ = $-18$ $T/ \mu_B$
\cite{38}. Interestingly, we find $A_{Zr} \approx A_{3d}$, which indicates that properties of the Zr magnetic sublattice in ZrZn$_{1.9}$ are close to properties of $3d$ hosts Fe,
Co, Ni. Using the calculated coupling constants we can estimate the magnetic moments for Zr and Zn: $\mu_{Zr}$ = 0.5 $\mu_B$/at. Zr and $\mu_{Zn}$  = 0.29 $\mu_B$/at. Zn.

Our estimation of magnetic moments and coupling constants is in contradiction with the measured net macroscopic moment of ZrZn$_2$ which according to previous studies lie in
the range 0.13 -- 0.23 $\mu_B$/f.u. The question on magnetic structure of ZrZn$_2$ then arises: obviously, it should be a complex one because the simple ferromagnetic order does
not fit our data. Notice that exactly the same question was posed by zero-field muon spin rotation experiments, which have provided evidence for an antiferromagnetic component
in the magnetic structure of ZrZn$_2$ \cite{23}. On the other hand, the zero-field magnetic structure of CeAl$_2$ which is isostructural to ZrZn$_2$, is reported to be non-chiral
spiral \cite{39}. Therefore, a spiral magnetic structure can also realize for Zr magnetic moments in ZrZn$_2$. The induced moments at Zn sites most likely order in a conical
magnetic structure because IMHF $B_{Zn}$ is at an angle 51-55$^{\circ}$ to the axis of easy magnetization and EFG [111] with magnetic moment aligned in the opposite
direction. This could result in the appearance of an induced Zn ferromagnetic component with a small macroscopic net magnetization, which at normal pressure smoothly decreases
as temperature increases up to $T_c$. The decrease is caused by the growth of the volume of the paramagnetic phase in the Zr sublattice, whose magnetic moments become
disordered.

\section{CONCLUSIONS}

We have performed accurate TDPAC-measurements of hyperfine magnetic field and electric field gradient on the $^{111}$Cd substitutional nuclei at Zr and Zn sites in ZrZn$_{1.9}$.
The TDPAC technique probing local properties offer more precise local atomic data on this compound.

From the TDPAC-data we have extracted values of magnetic moments at the Zr and Zn site. The magnetic moments are obtained from the coefficient of proportionality between the
HMF and atomic magnetic moment, and the value of HMF measured directly in TDPAC experiments. The coefficient of proportionality can be taken from (1) comparison between
HMF and local magnetic moments in other compounds, and (2) from {\it ab initio} electron band structure calculations \cite{36}. Both methods lead to consistent values ($\mu_
{Zr}$ = 0.5 $\mu_B$ and $\mu_{Zn}$ = 0.29 $\mu_B$) which are substantially larger (several times) than what has been believed so far.

This origin of this discrepancy within the present study remains unclear and requires a special consideration. One way to reconcile this finding with the macroscopic measurements
is to assume that the magnetic structure of the ground state is a long conical spiral where the magnetic moment of an individual unit cell corresponds to $ \sim 1 \mu_B$, while its
direction changes slowly from cell to cell so that the angle between the local and the global (net) magnetization is of the order of 10$^{\circ}$. Interestingly, our conclusion
that the magnetically ordered phase of ZrZn$_2$ is not a simple ferromagnet is also supported by the zero-field muon spin rotation ($\mu$SR) experiments. \cite{23}

Further studies of ZrZn$_2$ with methods that can selectively determine local magnetic properties are needed to determine the details of its magnetic order.

\acknowledgments

The authors are grateful to S.M. Stishov, V.B. Brudanin and N.G. Chechenin for support of this work. We are also grateful to I.I. Mazin for useful discussions and an opportunity
to use his unpublished data. The work was supported by the Russian Foundation for Basic Research (grant No. 14-02-00001) and by special programs of the Department of
Physical Science, Russian Academy of Sciences. The work at the Joint Institute for Nuclear Research was carried out under the auspices of a Polish representative in the JINR.

\end{document}